\documentstyle[floats,amssymb,aps,prl,epsfig]{revtex}

\begin{document}
\draft
\twocolumn[\hsize\textwidth\columnwidth\hsize\csname@twocolumnfalse\endcsname
\title{Unexpected Behavior of the Local Compressibility Near the B=0
Metal-Insulator Transition}
\author{S. Ilani, A. Yacoby, D. Mahalu and Hadas Shtrikman}
\address{Braun Center for Submicron Research, Dept. of Condensed Matter Physics,\\
Weizmann Institute of Science, Rehovot 76100, Israel.}
\maketitle
\begin{abstract}
We have measured the local electronic compressibility of a two-dimensional
hole gas as it crosses the B=0 Metal-Insulator Transition. In the metallic
phase, the compressibility follows the mean-field Hartree-Fock (HF) theory
and is found to be spatially homogeneous. In the insulating phase it
deviates by more than an order of magnitude from the HF predictions and is
spatially inhomogeneous. The crossover density between the two types of
behavior, agrees quantitatively with the transport critical density,
suggesting that the system undergoes a thermodynamic change at the
transition.
\end{abstract}
\pacs{PACS numbers: 71.30.+h, 73.40.-c, 05.70.Ce}
]

The recent discovery of a Two-Dimensional (2D) Metal-Insulator Transition
(MIT)\cite{kravchenko} has raised the fundamental question concerning the
existence of metallic 2D systems\cite{finkelstein,Abrahams}. In contrast to
the scaling theory of localization\cite{Abrahams}, which predicts that only
an insulating phase can exist in 2D, there is compelling evidence for
metallic-like behavior in a growing number of 2D systems\cite{mit,hanein98}.
To date, the vast majority of this evidence comes from transport
measurements. Clearly, if a true zero-temperature phase transition exists,
it may reveal itself also in the thermodynamic properties of the 2D gas,
such as the electronic compressibility, $\kappa ^{-1}=n^{2}\frac{\delta \mu
}{\delta n}$ (where $\mu $ is the chemical potential and $n$ is the carrier
density). For example, a crossover to a gapped insulator would result in a
vanishing compressibility at the transition\cite{si}, whereas a crossover to
an Anderson insulator would involve a continuous evolution of the
compressibility across the transition.

The compressibility or $\frac{\delta \mu }{\delta n}$ of a system reflects
how its chemical potential varies with density. For non-interacting
electrons it simply amounts to the single-particle Density of States (DOS),
which in 2D is density independent ($\frac{\delta \mu }{\delta n}=\frac{\pi
\hbar ^{2}}{m}$). This picture, however, changes drastically when
interactions are included. Exchange and correlation effects weaken the
repulsion between the electrons, thereby reducing the energy cost, thus
leading to negative and singular corrections to $\frac{\delta \mu }{\delta n}
$. Within the Hatree-Fock (HF) theory, which includes both the DOS and
exchange terms, one gets:
\begin{equation}
\frac{\delta \mu }{\delta n}=\frac{\pi \hbar ^{2}}{m}-\left( \frac{2}{\pi }%
\right) ^{\frac{1}{2}}\frac{e^{2}}{4\pi \varepsilon }n^{-1/2},  \label{HF}
\end{equation}
with the compressibility becoming negative at low enough densities.
Measurements of the macroscopic compressibility of 2D electron/hole gases,
in the metallic regime, have indeed confirmed this behavior\cite
{eisenstein94,shapira}.

In this work, we have expanded the study of $\mu (n)$ into the MIT regime.
Our measurements utilize several Single Electron Transistors (SETs),
situated directly above a Two-Dimensional Hole Gas (2DHG). This technique
allows us to determine the local behavior of $\mu (n)$ as well as its
spatial variations. Simultaneous macroscopic transport measurements were
conducted to ensure a precise determination of the MIT critical density in
the same sample. Our measurements indicate a clear thermodynamic change in $%
\mu (n)$ at the MIT. In addition, we find that the behavior of
$\mu (n)$ in the metallic phase follows the HF model and is
spatially homogeneous. The insulating phase, on the other hand, is
found to be spatially inhomogeneous. In this regime, $\frac{\delta
\mu }{\delta n}$ deviates by more than an order of magnitude from
the predictions of the HF model.

The samples are inverted GaAs/AlGaAs
heterostructures\cite{hanein97}. A layer of AlGaAs, $2000$\AA\
thick, separates the 2DHG from a $p^{+}$ layer
that serves as a back gate, and is separately contacted. A GaAs spacer, $500$%
\AA\ thick, separates the 2DHG from the doped GaAs layer situated just under
the surface (see inset in Fig. 1). The mobility of the 2DHGs ranges from $%
7\cdot 10^{4}\,cm^{2}/V\sec $ to $1.2\cdot 10^{5}\,cm^{2}/V\sec $ at $%
n=5\cdot 10^{10}\,cm^{-2}$ and $T=4.2\,K$. Biasing the back gate enables us
to vary the density of holes by two orders of magnitude, from $2\cdot
10^{11}\,cm^{-2}$ to $2\cdot 10^{9}\,cm^{-2}$. Our samples show a clear MIT.
A typical plot of the resistivity, $\rho $, vs. density at various
temperatures is shown in Fig. 1. A clear crossing point between metallic and
insulating behavior is observed. The critical density and resistivity,
determined from the low temperature crossing point, is $n_{c}\approx
2.1\cdot 10^{10}\,cm^{-2}$ and $\rho _{c}\approx 0.9\frac{h}{e^{2}}$.
Similar samples were previously used to study the transport properties of
the MIT in GaAs\cite{hanein98}.

A local measurement of $\mu $ requires the ability to measure
electrostatic potentials with high sensitivity and good spatial
resolution. It has been previously shown that both can be achieved
using an SET\cite{yoo,amir,wei}. We have therefore deposited
several aluminum SETs on top of an etched Hall bar mesa (inset in
Fig. 1). This configuration allows us to study simultaneously the
local thermodynamic behavior and the macroscopic transport
properties and, thereby, trace possible correlation between them.
Our technique relies on the fact that $\mu $ of the 2DHG changes
as its density is varied. At equilibrium, the Fermi energy is
constant across the sample, and therefore, a change in $\mu $
induces a change in the electrostatic potential, which is directly
measured by the SET. The size of the SET and its distance from the
2DHG determine its spatial resolution, estimated to be $0.1\times
0.5\,\mu m^{2}$ . The measured voltage sensitivity of the SET is
$10\,\mu V$. Similar technique has been previously employed to
image the local compressibility of a 2DEG in the quantum Hall
regime\cite{amir,wei}.

To demonstrate the strength of our technique, we first employ it
to study
the magnetic field ($B$) dependence of $\mu $ deep in the metallic regime ($%
n=1.96\cdot 10^{11}\,cm^{-2}\,$, see Fig. 2). Theoretically, the formation
of well-separated Landau levels at high magnetic fields is expected to
induce sharp saw-tooth like oscillations in $\mu $. The sharp drops are at
integer filling factors ($\nu $), and the slopes obey $\frac{\delta \mu }{%
\delta B}=\left( j+\frac{1}{2}\right) \frac{\hbar e}{m^{*}}\pm g^{*}\mu _{B}$%
, where $m^{*},\,g^{*}$ are the effective mass and g-factor, $j$ is the
Landau level number, and the $\pm $ relate to the two spin directions. We
find good quantitative agreement between this model and the measurement
using a single value for $m^{*}=0.34\cdot m_{e}$ and $g^{*}=2.7$. The
measured value of $m^{*}$ is also in good agreement with the known effective
mass of light holes in GaAs, $m^{*}=0.38\cdot m_{e}$\cite{eisenstein84}. The
position of the jumps allows us to determine the local density of the 2DHG
under the SET: $n_{l}=1.82\cdot 10^{11}\,cm^{-2}$, which deviates by only
7\% from the density measured macroscopically by transport techniques. This
small deviation demonstrates that the area beneath the SET is hardly
disturbed by its presence.

The main focus of this work is the density dependence of $\mu $ across the $%
B=0$ MIT. Both the theory and previous macroscopic measurements\cite
{eisenstein94,shapira} show that at low densities $\mu $ should increase
monotonically as $n$ is decreased (Eq. 1). Unexpectedly, our local
measurements demonstrate quite a different behavior. Several typical traces
of $\mu $ vs. $n$ are shown in Fig. 3. Fig. 3a shows the behavior in the
metallic regime and Fig. 3b shows the behavior across the MIT and in the
insulating regime. Instead of the expected monotonic dependence, $\mu $
exhibits a rich structure of oscillations. All the oscillations, including
the fine-structure seen on the left side of Fig. 3b, are completely
reproducible and do not depend on the measurement rate or sweeping
direction. Two distinct types of oscillations are observed: In the metallic
regime (Fig. 3a) we observe long saw-tooth oscillations with a typical
period in density of $(1-2)\cdot 10^{10}\,cm^{-2}$. Superimposed on them and
starting in close proximity to the MIT, a new set of rapid oscillations
emerges (Fig. 3b). Their typical period is an order of magnitude smaller: $%
(1-2)\cdot 10^{9}\,cm^{-2}$, and their amplitude grows continuously from
their point of appearance to lower densities. Similar behavior has been
measured on seven different SETs placed on five separate Hall bars from two
different wafers. Although the specific pattern of oscillations varies
between these Hall bars and changes also after thermal cycling, they all
have the same general characteristics. The qualitative change in the
oscillation pattern of $\mu (n)$ near the critical density suggests that the
system experiences a thermodynamic change at the MIT.

Out of the two types of oscillations, the easier ones to explain
are these on the metallic side. In Fig. 3a we compare the measured
oscillations with the expected monotonic behavior of the HF model.
In contrast to the model, the measured $\mu $ has a density
independent average which suggests that some kind of screening
mechanism is present in the system. While the negative slopes of
$\mu (n)$ (black symbols) follow the HF theory, there are apparent
additional drops (some of which are extremely sharp) between them
(gray symbols). This saw-tooth profile is reminiscent of the
behavior of the chemical potential of a quantum dot as a function
of its density\cite {Kouwenhoven} and suggests the existence of
discrete charging events in the dopant layer, situated between the
2DHG and the SET. Thus, the measured $\mu $ of the 2DHG varies
undisturbed along the negative slopes, until a certain bias is
built between the 2DHG and the SET that makes the charging of an
intermediate localized state energetically favorable. This causes
a screening charge to pop-in which result in a sharp drop in the
measured electrostatic potential, after which $\mu $ continues to
vary undisturbed until the next screening event occurs. Because
screening occurs only at discrete points we can reconstruct the
underlying $\mu (n)$ from the unscreened segments in the
measurement. We do this in Fig. 4, by collecting the slopes of
these segments from the saw-tooth profiles of five different SETs
placed on three separate Hall bars. Each single point in this
graph represent the slope of a well-defined segment, like the ones
emphasized in Fig 3a. In the metallic side all the data collapses
onto a single curve (the dashed line in Fig. 4 is a guide to the
eye). The prediction of the HF model is depicted by the solid line
in Fig. 4. The finite width of the 2DHG adds a positive
contribution to the compressibility as shown by the dotted line.
It is evident that both theories describe the data to within a
factor of two over more than an order of magnitude in density,
throughout the metallic regime. The fact that different Hall bars
and different SETs (namely different locations within the same
Hall bar) produce the same dependence, suggests that the metallic
phase is{\em \ homogeneous in space}.

The sudden appearance of rapid oscillations as the system crosses
to the insulating side already signifies that something dramatic
happens at the transition. Assuming that this new set of
oscillations is caused by the same mechanism, namely screening by
traps, we can proceed and extract its slopes and add them to the
same plot of $\delta \mu /\delta n$ (see Fig. 4). Unlike in the
metallic phase where the system clearly has a negative $\delta \mu
/\delta n$, namely negative compressibility, in the insulating
phase we a-priory do not know the sign of the compressibility. We,
therefore, plot in the left side of Fig. 4 both negative and
positive slopes, all of which deviate considerably from the
expected $n^{-1/2}$ power law. The deviation becomes greater than
an order of magnitude at our lowest density\cite {comment}. We see
that $\delta \mu /\delta n$ starts deviating from the theoretical
curve in close proximity to the transport-measured critical
density, signifying a change in the screening properties of the
2DHG at the transition. Another intriguing result is the
non-universal behavior of the slopes on the insulating side. It
should be emphasized that each slope is determined from a complete
segment in the $\mu \left( n\right) $ curve (see inset to Fig. 3b)
and is, therefore, determined very accurately. The fluctuations in
the slopes, seen even in a measurement done using a single SET,
are completely reproducible and suggest that mesoscopic effects
are present. Furthermore, the average behavior of $\delta \mu
/\delta n$ in this fluctuating regime is seen to be position
dependent. An example of that is shown in the inset in Fig. 4
where we plot $\delta \mu /\delta n$ measured by two separate SETs
on the same Hall bar. Such dependence on position
indicates that once the system crosses into the insulating phase it {\em %
ceases} {\em to be spatially homogeneous}.

The charge traps have a clear signature in our measurements, emphasizing
their important role in the thermodynamic ground state of the system. It was
recently suggested by Altshuler and Maslov\cite{altshuler} that a gas that
is in electro-chemical equilibrium with charge traps might show
metallic-like behavior of the resistivity. This behavior is a result of a
temperature-dependent scattering mechanism\cite{altshulerA} evoked by the
coupling to the trap system. Although it is clear from our measurements that
charging of the traps takes place as the density of holes is varied, we are
currently unable to determine experimentally whether this coupling is
responsible for the metallic behavior seen in our samples.

The large values of $\delta \mu /\delta n$ in the insulating phase is
intriguing. Currently there are no theoretical predictions for large
negative $\delta \mu /\delta n$, which in turn leads to a sharp rise in $\mu
$ as the density is reduced. However, we would like to dwell on several
scenarios in which the measured $\delta \mu /\delta n$ becomes {\em positive}
at the transition and increasingly large as the density is further reduced.
The first scenario assumes the formation of voids in the 2DHG. Such voids
are unable to screen the back gate voltage and, hence, would result in an
added large and positive contributions to $\delta \mu /\delta n$. In this
scenario, the size and shape of the voids, as function of decreasing
density, are responsible for the large and seemingly random set of positive
slopes observed. A second scenario involves the formation of isolated
puddles in the 2DHG\cite{Shimshoni}. These puddles screen the back gate
discretely and, hence, would produce large and positive slopes of $\mu
\left( n\right) $. It should be noted that both scenarios suggest a
percolative like transition\cite{he,meir} and are in accord with our
observation that the 2DHG is inhomogeneous in the insulating phase.

In conclusion, we have measured the dependence of $\mu (n)$ across the MIT.
Our measurements clearly indicate a thermodynamic change in the screening
properties of the 2DHG at the transition. We find the metallic phase to be
spatially homogenous and to behave according to the predictions of the HF
model. The insulating phase, on the other hand, deviates significantly from
the HF predictions and is spatially inhomogeneous.

We benefited greatly from discussions with A. M. Finkel'stein, Y. Hanein, Y.
Meir, D. Shahar, A. Stern and C. M. Varma. This work was supported by the
MINERVA foundation, Germany.

\begin{figure}
\caption{Density dependence of the resistivity: Traces from top to
bottom on the right correspond to T=600, 550, 500, 450, 400, 300
and 250 mK . The low-T crossing point is at $n_{c}=2.1\cdot
10^{10}\,cm^{-2}$. Inset: An SEM micrograph of the SET, placed on
top of a schematic structure along with the measurement circuit.}
\label{fig1}
\end{figure}
\begin{figure}
\caption{Measured B-dependence of $\mu $. The solid line is the
predicted
Landau fan with $m^{*}=0.34\cdot m_{e}$ and $g^{*}=2.7$. Oscillations in $%
\mu $ are observable up to $\nu =7$ . The {\em local density}
extracted from the magnetic field values at integer $\nu $ is
$n_{l}=1.82\cdot 10^{11}\,cm^{-2}$ which deviates only by 7\% from
the transport-measured macroscopic density, $n=1.96\cdot
10^{11}\,cm^{-2}$.} \label{fig2}
\end{figure}
\begin{figure}
\caption{(a) $\mu (n)$ in the metallic regime. The HF predictions
(\ref{HF}) are depicted by the solid line. The negative-slopes are
high lighted (dark symbols) to demonstrate their resemblance to
the HF model. (b) $\mu (n)$ across the MIT and in the insulating
region. Inset: A closer look at the data in the insulating regime.
Each slope is composed of many data points allowing for an
accurate determination of the slopes.} \label{fig3}
\end{figure}
\begin{figure}
\caption{$\delta \mu /\delta n$ collected from five SETs on three
different hall bars from two different wafers. In the insulating
regime, both negative and positive slopes are shown (closed and
open symbols respectively). Each point corresponds to a
well-defined segment in the $\mu (n)$ trace. The point marked by
an arrow corresponds to the marked segment in Fig. 3a. Inset:
Results from two SETs on the same device ($\bullet ,\blacktriangle
$) demonstrating the spatial dependence of $\delta \mu /\delta n$
on the insulating side.} \label{fig4}
\end{figure}


\begin{references}
\bibitem{comment}  When the compressibility is very small, the linear
relation between the back gate voltage and $n$ becomes inaccurate.
Correcting for this (see Ref. \cite{eisenstein94}) increases the discrepancy
between the data and the model.
\end{references}
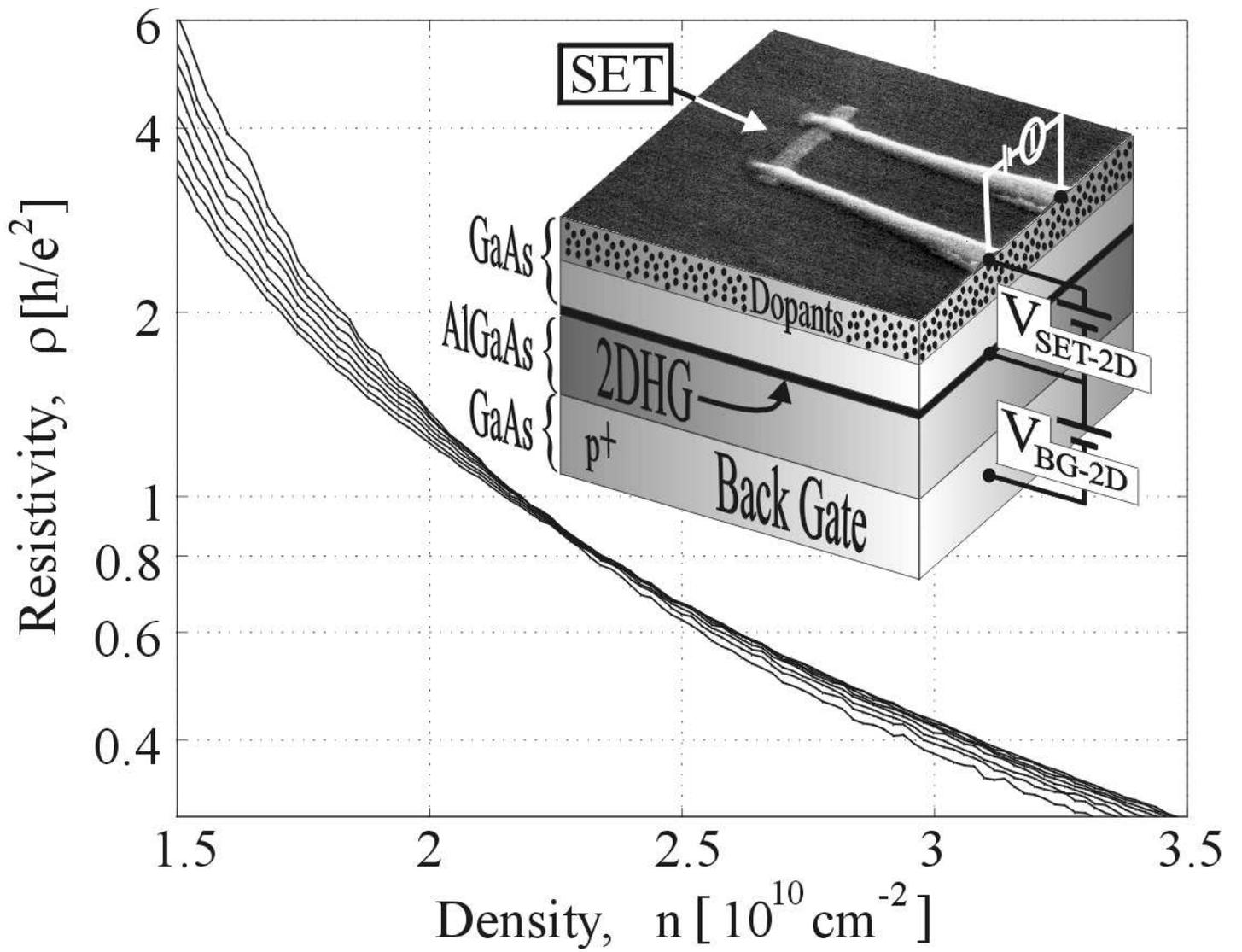
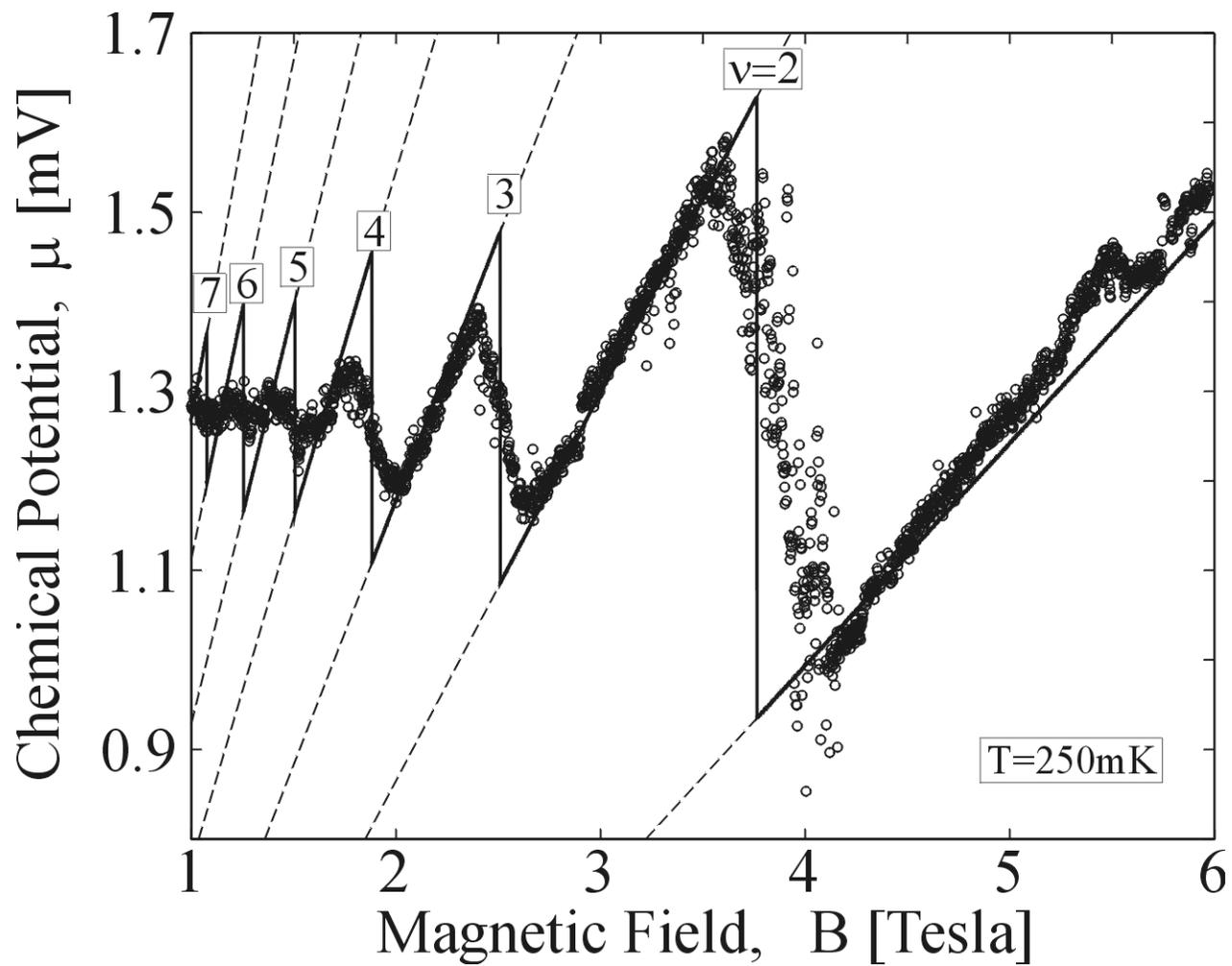
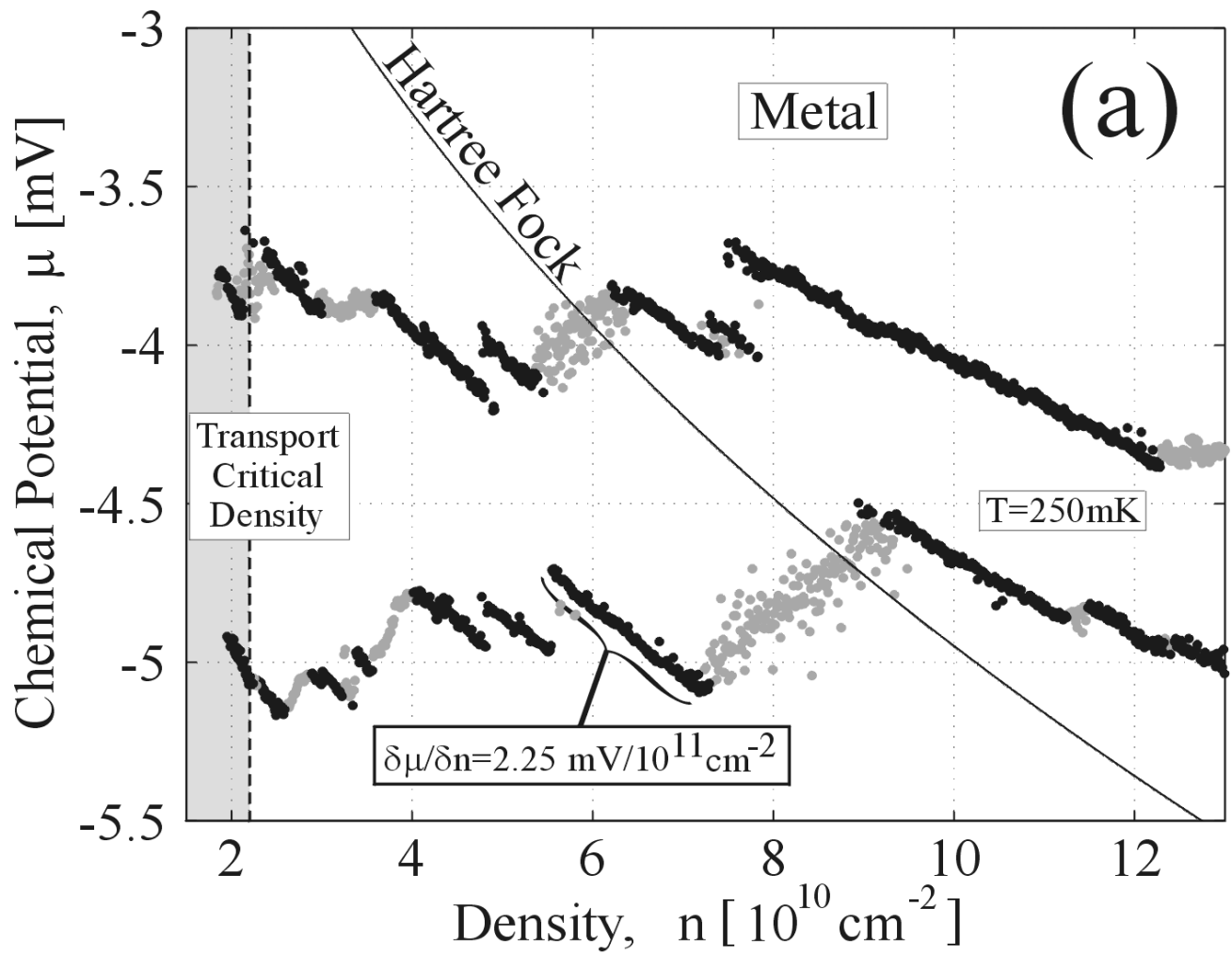
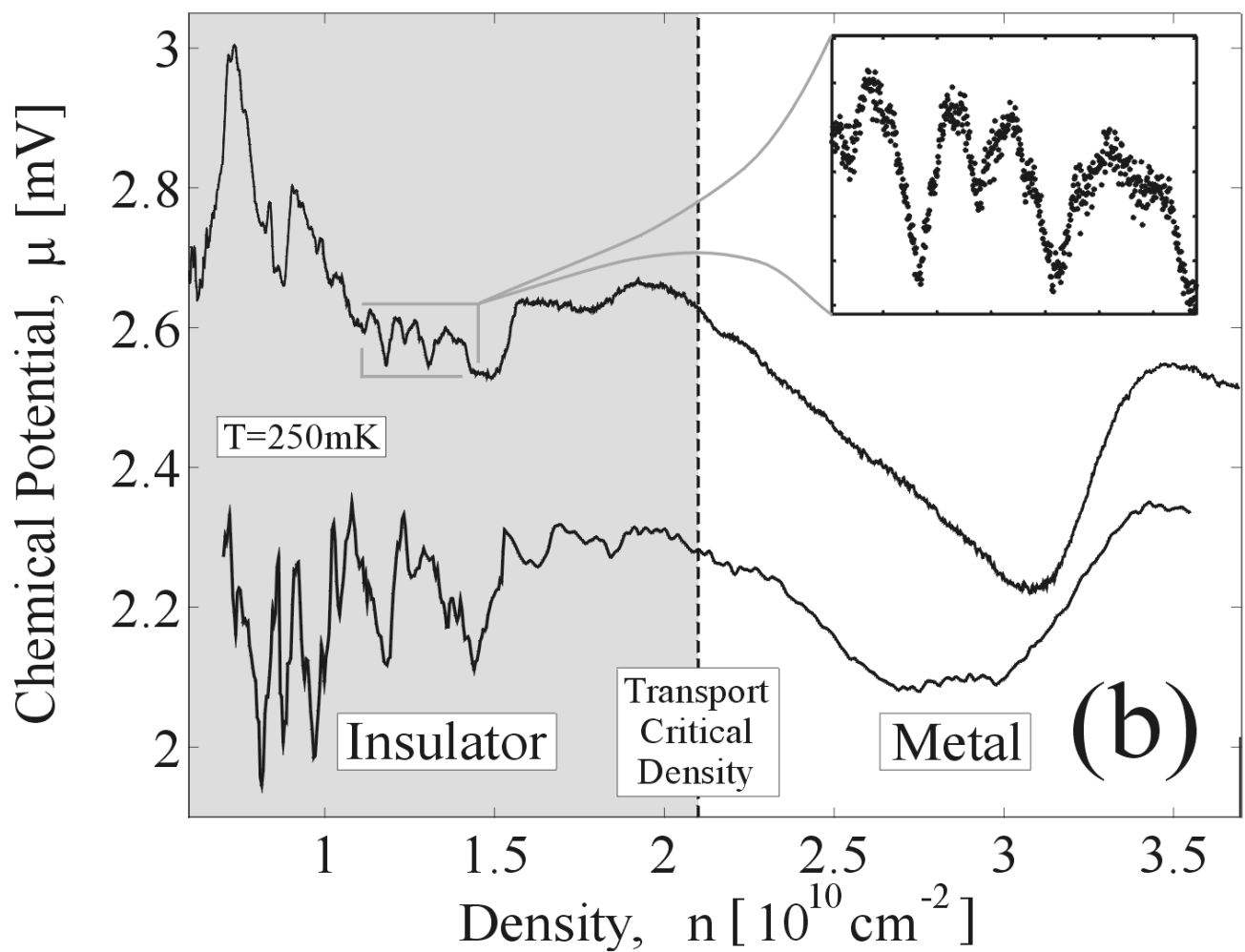
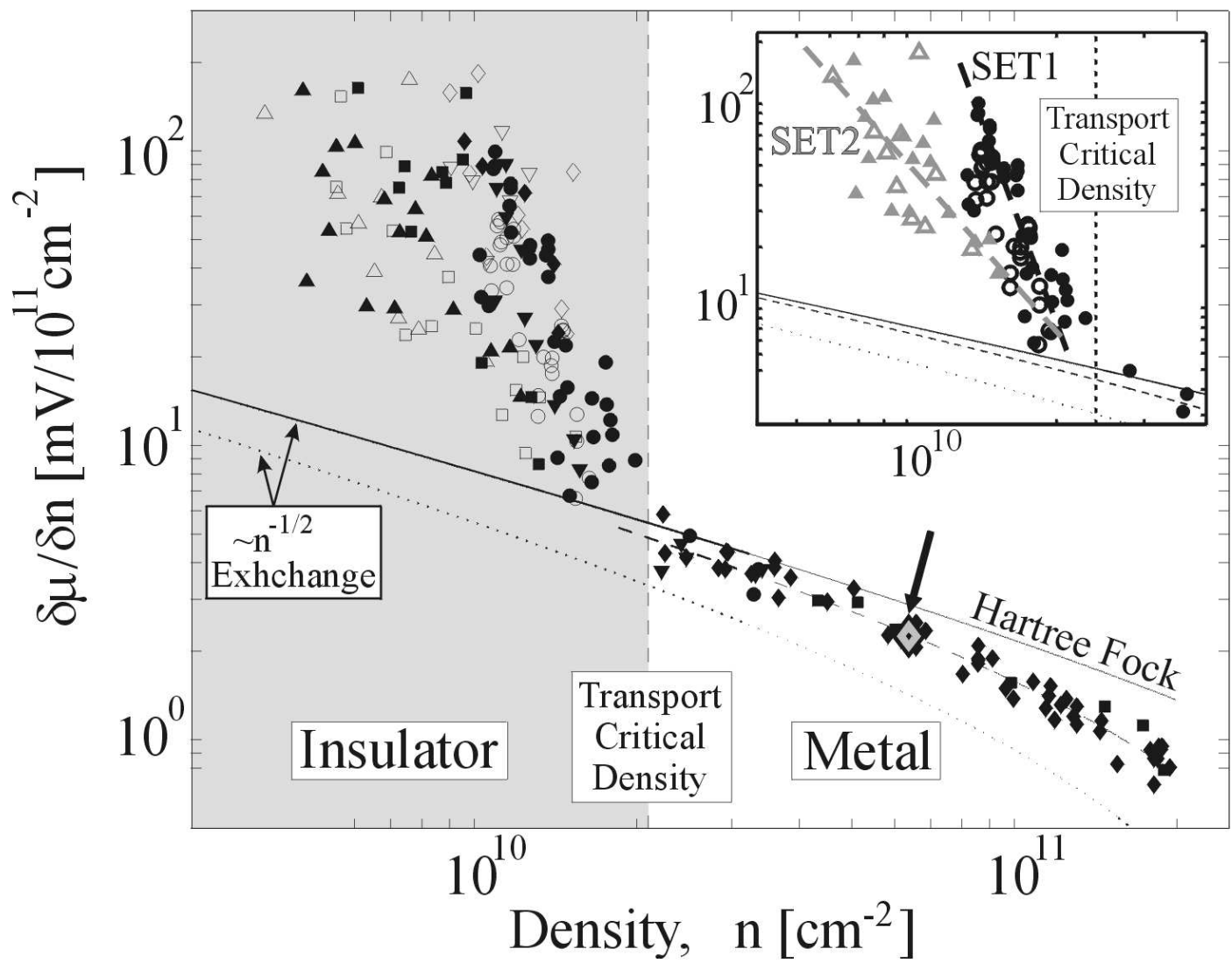
\end{document}